\documentclass[aps,twocolumn,groupedaddress]{revtex4-1}

\usepackage{amsmath,bm}
\usepackage{graphicx}
\usepackage[caption=false]{subfig}
\usepackage{bookman}
\usepackage{times}
\usepackage{ctable}
\usepackage{multirow}
\usepackage{floatrow}
\usepackage{wrapfig}
\usepackage{hyperref}
\usepackage{SIunits}

\begin{document}




\title{EMC/FDTD/MD simulation of carrier transport and electrodynamics in two-dimensional electron systems}


\author{N.~Sule}\email{sule@wisc.edu}
\author{K.~J.~Willis}
\author{S.~C.~Hagness}
\author{I.~Knezevic}\email{knezevic@engr.wisc.edu}

\affiliation{Department of Electrical and Computer Engineering, University of Wisconsin-Madison, Madison, WI 53706, USA}

\begin{abstract}

We present the implementation and application of a multiphysics simulation technique to carrier dynamics under electromagnetic excitation in supported two-dimensional electronic systems. The technique combines ensemble Monte Carlo (EMC) for carrier transport with finite-difference time-domain (FDTD) for electrodynamics and molecular dynamics (MD) for short-range Coulomb interactions among particles. We demonstrate the use of this EMC/FDTD/MD technique by calculating the room-temperature \emph{dc} and \emph{ac} conductivity of graphene supported on $\rm{SiO_2}$.

\end{abstract}
\maketitle

\section{Introduction}
\label{intro}

Electronic properties of supported two-dimensional (2D) materials, such as the single layer graphene \cite{Das-Sarma:2011fk} or $\rm{MoS_2}$ \cite{Fontana:2013uq}, and quasi-2D materials, such as semiconductor membranes \cite{Zhang:2006uq}, have generated a lot of interest in recent years. These 2D electronic systems (2DESs) have potential applications as electronic \cite{Lin05022010,B.:2011vn} and optoelectronic devices \cite{Bonaccorso:2010ys,doi:10.1021/nn2024557}, THz detectors \cite{Ju:2011kx}, as well as chemical and biologicial sensors \cite{C1CS15270J}. Realizing these applications requires an understanding of carrier transport in 2D materials in the presence of electromagnetic fields, while accounting for the strong influence of the supporting substrate \cite{sonde:132101,PhysRevB.77.195415} and the impurities found near the 2DES/substrate interface \cite{Adam20112007,PhysRevLett.98.186806}.

A multiphysics numerical solver that combines ensemble Monte Carlo (EMC) simulation of carrier transport with the finite-difference time-domain (FDTD) technique for solving Maxwell's curl equations and molecular dynamics (MD) for short-range particle interactions \cite{willis:063714} can provide insight into the carrier transport and electrodynamics of 2DESs. Unlike most device simulation tools that implement EMC coupled with a quasi-electrostatic solver of Poisson's equation \cite{tomizawa1993numerical,jacoboni1989monte}, EMC/FDTD/MD couples EMC with a fully electrodynamic solver \cite{xaldgh,zskugadf,anuayw,willis:062106}, which enables simulation of carrier dynamics under electromagnetic excitation, from low frequencies (including \textit{dc}) to the THz frequency range. (At frequencies above THz, interband transitions in semiconductors become important and the classical view of carrier--field interaction is no longer sufficient.)

Grid-based solvers, such as Poisson solvers or FDTD, cannot accurately capture the forces among charges on spatial scales smaller than the size of a grid cell. In order to include short-range (sub-grid cell)  interactions, EMC-FDTD has been extended through coupling with MD \cite{PhysRevLett.56.1295,Ferry1991119,822288,Vasileska:2008:1546-1955:1793}. The EMC/FDTD/MD technique includes accurate pair-wise, short-range, real-space Coulomb forces among carriers and between carriers and charged impurities, together with the full electrodynamics solution for long-range Coulomb fields \cite{willis:063714}. EMC/FDTD/MD has been used to accurately calculate the conductivity of bulk Si in the THz frequency range, where the usual Drude model fails, with good agreement to experimental data \cite{willis:063714,willis:122113}.

\begin{figure}
\centering
\includegraphics[width=\columnwidth]{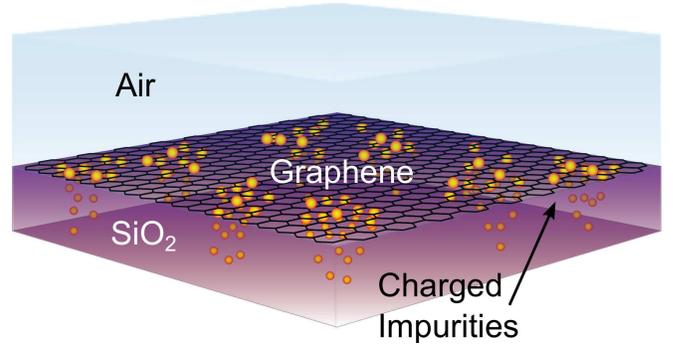}
\caption{Schematic of the simulated structure: single-layer graphene rests on an $\rm{SiO_2}$ substrate, with charged impurities present near the interface between the two. Carrier transport is simulated by including both electrons and holes in the graphene layer, while the positively charged ions near the interface and within the $\rm{SiO_2}$ substrate remain stationary.}
\label{figsimstructure}
\end{figure}

In this paper, we describe the EMC/FDTD/MD technique as applied to simulating carrier transport in 2DESs. We simulate a structure with a single graphene layer resting on top of an $\rm{SiO_2}$ substrate, with impurity ions near the interface (Fig.~\ref{figsimstructure}). We describe the constituent techniques (Sec. \ref{sec:1}) and the procedure for coupling them (Sec. \ref{sec:2}). As the distribution of impurity ions is important for the overall carrier transport properties \cite{PhysRevLett.107.156601}, we describe the generation of a desired impurity distribution, from uniformly random to clustered (spatially correlated). In Sec. \ref{sec:4}, we give examples of \emph{dc} and \emph{ac} conductivity of supported graphene in the presence of charged impurities, as calculated using EMC/FDTD/MD. We conclude with Sec. \ref{sec:5}.

\section{Constituent Techniques}
\label{sec:1}
This section provides a brief overview of the constituent techniques of the combined EMC/FDTD/MD solver, with focus on the implementation details relevant for the simulation of carrier transport in graphene.

\subsection{Ensemble Monte Carlo (EMC)}
\label{subsec1:1}
Ensemble Monte Carlo simulates carrier dynamics in the diffusive transport regime \cite{jacoboni1989monte}. This method yields a solution to the Boltzmann transport equation by using statistically appropriate stochastic sampling of the relevant relaxation mechanisms, free-flight times, and angular distributions of momenta \cite{RevModPhys.55.645}. In an EMC simulation, the evolution of a large ensemble of carriers [typically $O(10^5)$] is tracked over time. The evolution of physical properties of interest, such as the carrier average drift velocity or kinetic energy, are calculated by averaging over the ensemble.

During the simulation, each carrier undergoes periods of "free flight", or drift, under the influence of the Lorentz force,
\begin{equation}
\vec{F}=q(\vec{E}+\vec{v}\times\vec{B}),\label{eqnLorentzforce}
\end{equation}
interrupted by instantaneous scattering events. In Eq.~(\ref{eqnLorentzforce}), $\vec{E}$ and  $\vec{B}$ are the electric and magnetic field vectors, respectively, $q$ is the carrier charge, and  $\vec{v}$ is the carrier velocity. In our EMC simulation, we include both electrons and holes in graphene. The carrier velocity in graphene is given by $\vec{v}=v_\mathrm{F}\frac{\vec{k}}{|\vec{k}|}$, where $v_\mathrm{F}$ is the Fermi velocity and $\vec{k}$ is the carrier momentum. For a free flight of duration $t_\mathrm{d}$ (obtained stochastically \cite{RevModPhys.55.645}), the momentum and energy of a carrier are updated based on the Lorentz force and $E\text{-}k$ dispersion, as
\begin{subequations}
\begin{align}
\vec{r}_{\mathrm{new}}&=\vec{r}_{\mathrm{old}}+\int_0^{t_{\mathrm{d}}}\vec{v}(t)\,dt\, ,\label{eqnposition}\\
\vec{k}_{\mathrm{new}}&=\vec{k}_{\mathrm{old}}+\hbar^{-1}\int_0^{t_{\mathrm{d}}} \vec{F} (\vec{r}(t))\,dt\,,\label{eqnmotion}\\
E&=\hbar v_\mathrm{F}|\vec{k}_{\mathrm{new}}|.\label{eqndispersion}
\end{align}
\end{subequations}

\begin{figure}
\centering
\includegraphics[width=\columnwidth]{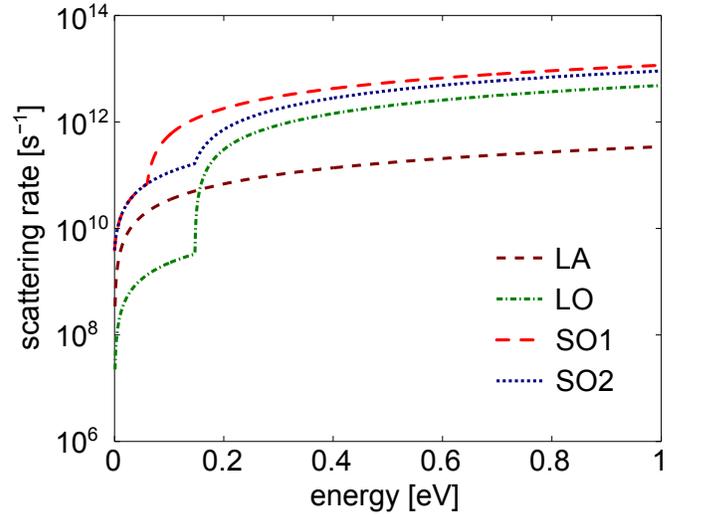}
\caption{Electron--phonon scattering rates in graphene, calculated using the third-nearest-neighbor tight-binding Bloch wave functions (3NN TBBW) \cite{sule:053702}. Scattering due to longitudinal acoustic (LA) and optical (LO) phonons intrinsic to graphene, as well as the surface optical phonons (SO1 and SO2) between $\rm{SiO_2}$ and graphene, is included.}
\label{figscatrates}
\end{figure}

\noindent The electron-phonon scattering rates in graphene are calculated based on the third-nearest-neighbors tight-binding Bloch wave functions (3NN TBBW) \cite{sule:053702} and are shown in Fig.~\ref{figscatrates}. (Near the Dirac point, electron and hole dispersions, as well as their rates for scattering with phonons, are considered to be identical.) The deformation potential constants ($D_{\mathrm{ac}}= 12$ eV and $D_{\mathrm{op}}= 5\times 10^{11}$ eV/m) were determined based on fitting the longitudinal acoustic (LA) and optical (LO) phonon scattering rates to the rates calculated from first principles \cite{PhysRevB.81.121412}. The surface optical (SO1 and SO2) phonon scattering rates are calculated from the interaction Hamiltonian following the dielectric continuum model of surface phonons \cite{PhysRevB.82.115452}.

\subsection{Finite-Difference Time-Domain (FDTD) method}
\label{subsec1:2}
In FDTD \cite{taflove2005computational}, the time-dependent Maxwell's curl equations
\begin{subequations}\label{eqMaxwellBoth}
\begin{align}
\mu\frac{\partial\vec{H}}{\partial t} &= -\nabla\times\vec{E}-\vec{M}\label{eqnMaxwellH}\\
\epsilon\frac{\partial\vec{E}}{\partial t} &= \nabla\times\vec{H}-\vec{J}\label{eqnMaxwellE}
\end{align}
\end{subequations}
are discretized using a centered-difference scheme for the partial derivatives in both space and time \cite{1138693}. The field components ($\vec{E}$ and $\vec{H}=\mu^{-1}\vec{B}$) are spatially staggered. Equations (\ref{eqMaxwellBoth}) are solved by leapfrog time integration: the $\vec{E}$-field and $\vec{H}$-field updates are offset by half a time step, yielding a fully explicit scheme with second-order accuracy in time.

\begin{figure}
\centering
\includegraphics[width=\columnwidth]{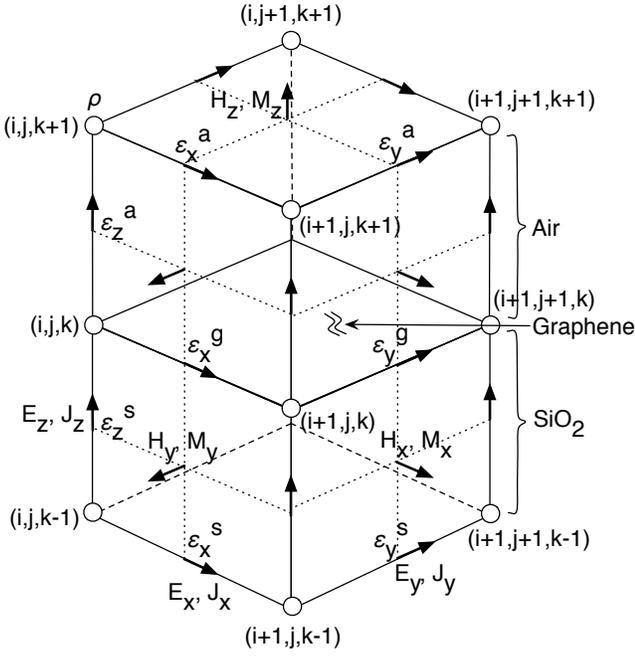}
\caption{Schematic of two FDTD grid cells at the air/graphene/$\rm{SiO_2}$ interface. The top cell is assumed to be filled with air and the bottom cell is assumed to be filled with $\rm{SiO_2}$. The central plane between these two cells represents the graphene layer. The FDTD field and current vectors ($\vec{E}$, $\vec{J}$, $\vec{H}$, $\vec{M}$), staggered in space, are shown with arrows on the grid edges and grid faces. The charge density ($\rho$) is defined on the grid cell corners shown with open circles.}
\label{figfdtdgrid}
\end{figure}

Fig.~\ref{figfdtdgrid} shows a schematic of two FDTD grid cells above and below the plane of graphene. The field components ($\vec{E}$ and $\vec{H}=\mu^{-1}\vec{B}$) are spatially staggered (Fig.~\ref{figfdtdgrid}). The $E_x$ and $E_y$ field components in the $(k+1)$-th plane, as well as the $E_z$ field components in the $(k+1/2)$-th plane, are updated assuming material properties corresponding to air $(\epsilon_\mathrm{a}=1)$; the $E_x$ and $E_y$ field components in the $(k)$-th plane are updated assuming graphene properties $(\epsilon_\mathrm{g}=2.45)$; and the $E_x$ and $E_y$ field components in the $(k-1)$-th plane, as well as the $E_z$ field components in the $(k-1/2)$-th plane, are updated assuming $\rm{SiO_2}$ properties $(\epsilon_\mathrm{s}=3.9)$.

The convolutional perfectly matched layer (CPML) absorbing boundary condition \cite{MOP:MOP14}, with a thickness of $10\text{--}20$ grid cells, is applied at the top and bottom horizontal boundaries of the domain. We use periodic boundary conditions for the four vertical boundary planes perpendicular to the graphene sheet. An incident plane wave is introduced via the total-field scattered field (TFSF) framework \cite{taflove2005computational}: electric and magnetic currents ($\vec{J}$ and $\vec{M}$) are calculated using the surface equivalence and applied at the boundary between the total-field and scattered-field regions in order to source a propagating plane wave. For \emph{dc} excitation, the electric field component is forced to remain constant once the peak magnitude of the plane wave is attained.

\subsection{Molecular Dynamics (MD)}
\label{subsec1:3}
Molecular dynamics simulates short-range interactions in classical many-particle systems \cite{rapaport2004art}. For a collection of electrons, holes, and charged ions, the particle-particle short-range interactions we include are the direct and exchange Coulomb forces among carriers (electrons and holes), and the direct Coulomb forces between carriers and ions \cite{willis:063714}. We only calculate the pairwise interactions among the particles present within a $3\times 3\times 3$-cell volume of one another, in order to minimize the computational burden in MD, which scales as $N^2$, $N$ being the number of interacting particles \cite{alder:459}.

The carriers in MD are described by Gaussian wave packets \cite{PhysRevLett.65.1619,joshi:2369} with a finite size $r_\mathrm{c}$, corresponding to the effective radius of the Hartree-Fock exchange-correlation hole \cite{PhysRevB.61.7353,willis:063714,willis:122113}
\begin{equation}
\phi_{\mathrm{\vec{k}}_\mathrm{i}}(\vec{r}_\mathrm{i})=\frac{1}{(2\pi r_c)^{3/4}}\exp{\left[-\frac{\vec{r}_\mathrm{i}^2}{4r^2_\mathrm{c}}+\vec{k}_\mathrm{i}\cdot\vec{r}_\mathrm{i}\right]}
\end{equation}
where $\vec{k}_\mathrm{i}$ and $\vec{r}_\mathrm{i}$ are the wave vector and position of the $i$-th carrier, respectively. The charged impurity ions are also described by a Gaussian profile with a characteristic radius of $r_\mathrm{d}=3.5$ \AA. Considering that the type and charge of impurity ions appear to be strongly dependent on processing \cite{LiangACSNano}, we considered positive ions with a unit charge. We swept \emph{dc} conductivity as a function of $r_\mathrm{d}$ and picked an $r_\mathrm{d}$ value from a range over which the \emph{dc} conductivity does not significantly vary with $r_\mathrm{d}$. These Gaussian profiles for charges in MD avoid large unphysical forces between pointlike particles that can lead to instability and errors \cite{PhysRevB.61.7353}. The Coulomb forces between particles with such Gaussian profiles are given below \cite{willis:063714}:
\begin{subequations}\label{eqnMDforce}
\begin{align}
\vec{F}_{\mathrm{ij}}^{\mathrm{D,d}} &=-\frac{q_\mathrm{i}Q_\mathrm{j}}{4\pi\epsilon_\mathrm{g}}\nabla_{\vec{r}_\mathrm{i}}\left[\frac{1}{r_{\mathrm{ij}}}\mathrm{erf}\left(\frac{r_{\mathrm{ij}}}{2r_\mathrm{d}}\right)\right], \label{eqnMDdirection}\\
\vec{F}_{\mathrm{ij}}^{\mathrm{D,c}} &=-\frac{q_\mathrm{i}q_\mathrm{j}}{4\pi\epsilon_\mathrm{g}}\nabla_{\vec{r}_\mathrm{i}}\left[\frac{1}{r_{\mathrm{ij}}}\mathrm{erf}\left(\frac{r_{\mathrm{ij}}}{2r_\mathrm{c}}\right)\right], \label{eqnMDdirectcar}\\
\vec{F}_{\mathrm{ij}}^{\mathrm{Ex}} &=-\frac{q^2}{8\pi^{3/2}\epsilon_\mathrm{g} r^4_{\mathrm{c}}}\frac{\vec{r}_{\mathrm{ij}}}{k_{\mathrm{ij}}}\label{eqnMDexch1}\\
&\times\exp{\left(-\frac{r^2_{\mathrm{ij}}}{4r^2_\mathrm{c}}-k^2_{\mathrm{ij}}r^2_\mathrm{c}\right)}\int_0^{k_{\mathrm{ij}}r_\mathrm{c}}\mathrm{d}t\,\mathrm{e}^{t^2},\nonumber\\
\vec{G}_{\mathrm{ij}} &= -\frac{q^2}{4\pi^{3/2}\epsilon_\mathrm{g} r^2_\mathrm{c}\hbar}\label{eqnMDexch2}\\
&\times \nabla_{\vec{k}_{\mathrm{ij}}}\left[\frac{1}{k_{\mathrm{ij}}}\exp{\left(-\frac{r^2_{\mathrm{ij}}}
{4r^2_\mathrm{c}}-k^2_{\mathrm{ij}}r^2_\mathrm{c}\right)}\int_0^{k_{\mathrm{ij}}r_\mathrm{c}}\mathrm{d}t\,\mathrm{e}^{t^2}\right].\nonumber
\end{align}
\end{subequations}
In the above equations, $\vec{F}_{\mathrm{ij}}^{\mathrm{D,d}}$ is the direct Coulomb force between the $i$-th carrier and the $j$-th ion, while $\vec{F}_{\mathrm{ij}}^{\mathrm{D,c}}$ is the direct force between the $i$-th and $j$-th carriers. $\vec{F}_{\mathrm{ij}}^{\mathrm{Ex}}$ is the ``exchange force'' between the $i$-th and $j$-th carriers having the same charge and spin, and $\vec{G}_{\mathrm{ij}}$ is a small correction to the velocity of the $i$-th carrier due to the $j$-th carrier, stemming from the exchange interaction. Also, $\vec{r}_{\mathrm{ij}}=\vec{r}_\mathrm{i}-\vec{r}_\mathrm{j}$, $\vec{k}_{\mathrm{ij}}=\vec{k}_\mathrm{i}-\vec{k}_\mathrm{j}$, erf($x$) denotes the error function, $\epsilon_\mathrm{g}$ is the relative permittivity of graphene, while $q$ and $Q$ are the carrier and impurity charge, respectively. These forces, given in Eq.~(\ref{eqnMDforce}), are calculated numerically at the beginning of the simulation for a small fixed volume in the real and momentum spaces, and stored in lookup tables.

\section{Coupled EMC/FDTD/MD}
\label{sec:2}
At the beginning of the coupled simulation, the carrier ensemble is initialized based on the equilibrium Maxwell-Boltzmann distribution. Poisson's equation is solved to calculate the initial microscopic field distribution stemming from all the charges in the domain (electrons, holes, and charged impurities). As time-stepping commences, carriers in the EMC module drift under the action of the fields and scatter according to the appropriate scattering mechanisms. Carrier motion results in a current density that can be calculated from carrier velocities. The positions of the carriers also change, leading to different short-range electrostatic interactions. Thus, the current densities and the new carrier positions can now be used to adjust the fields acting on the carriers (in the FDTD and MD modules). We ensure that the fields from FDTD and MD are not double counted in the vicinity of the charges \cite{willis:063714}, by subtracting the grid-based (FDTD) contribution of fields in the vicinity of the charges from the total pair-wise MD contribution of the fields. Moreover, to correctly represent fields arising from non-uniform and time-varying charge densities in FDTD, the initial field distribution must satisfy Gauss's law and the continuity equation must be enforced at each time-step \cite{willis:063714}. Thus, accurate and stable coupling of the EMC, FDTD, and MD methods requires proper initialization and assignment of charges to the grid, initialization of the fields including the grid-based fields of the impurity ion distribution, calculation of the current density, and calculation of the MD fields for updated positions. In the following subsections, we describe the four requirements for coupling in further detail.

\subsection{Charge initialization and assignment}
\label{subsec2:1}

\subsubsection{Initialization}
We assume that the Fermi level and charge density in graphene can be modulated by a back gate, located at the bottom of the SiO$_2$ substrate. The simulation domain is not charge neutral due to the assumption of a back gate, which is unlike the previous applications of the EMC/FDTD/MD method \cite{willis:063714,willis:122113}. For a given Fermi level and temperature, the density of carriers (electrons and holes) in graphene is given by \cite{fang:092109}
\begin{equation}
n=\frac{\pi}{6}\left(\frac{kT}{\hbar v_\mathrm{F}}\right)^2\frac{\int\mathrm{d}u\,u[1+\exp{(u\mp\eta)]^{-1}}}{\int\mathrm{d}u\,u[1+\exp(u)]^{-1}},\label{eqncardensity}
\end{equation}
where $u=\frac{E}{k_\mathrm{B}T}$ and $\eta=\frac{E_\mathrm{F}}{k_\mathrm{B}T}$. Here, the minus (plus) sign corresponds to electron (hole) density. The size of the simulation domain is chosen such that the total number of carriers is $O(10^5)$; molecular dynamics calculation necessitates that one numerical particle correspond to one physical particle \cite{willis:063714}. The momentum and energy of the carriers are defined according to the Maxwell-Boltzmann distribution. The carriers are initially positioned according to a uniform random distribution throughout the graphene plane.

\begin{figure}
\centering
\includegraphics[width=\columnwidth]{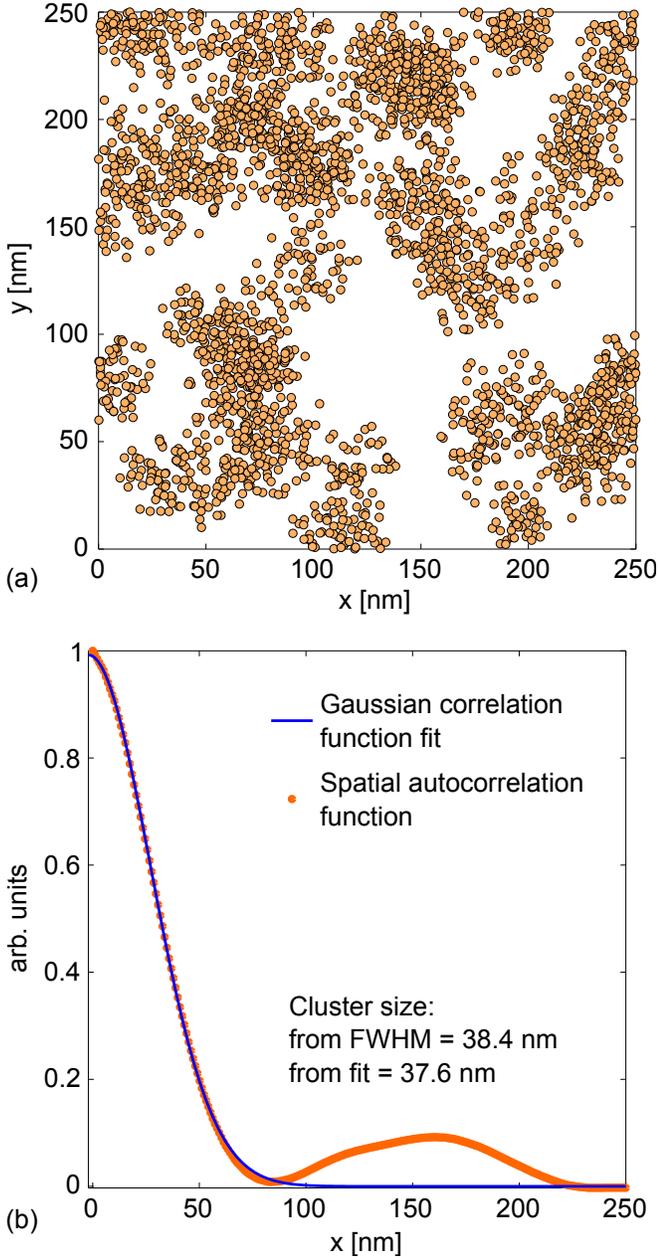}
\caption{(a) Example of a numerically generated clustered impurity distribution with a sheet density of $5\times10^{11}\usk\centi\meter^{-2}$. (b) The spatial autocorrelation function (SACF) of the numerically generated distribution (orange circles) is fitted with a Gaussian correlation function (blue line) of the form $\exp(-r^2/\lambda^2)$, where $\lambda$ is the correlation length. The average cluster size/correlation length estimated from the full width at half maximum (FWHM) of the SACF is $38.4\usk\nano\meter$, while that estimated from a Gaussian fit is $37.6\usk\nano\meter$.}
\label{figimpdist}
\end{figure}

The impurity ions are distributed below the graphene plane, down to a depth of $10\usk\nano\meter$. In our tests, we have observed that charged impurities, for reasonable sheet densities ($<10^{12}\usk\centi\meter^{-2}$), do not appreciably affect transport in the graphene layer beyond a depth of about $10\usk\nano\meter$. The type and charge of the relevant impurities vary with the processing details \cite{LiangACSNano}; for simplicity, here we use positive impurity ions with unit charge. In the literature, density of impurities is typically described via a cumulative sheet density, in units of $\centi\meter^{-2}$, however, these ions are distributed throughout the three-dimensional substrate. For a generated 3D distribution of ions, the sheet density is obtained by integrating over a depth equal to $2r_\mathrm{d}$ (i.e. twice the typical ion radius) and averaging over the $10\usk\nano\meter$ depth. The positions of impurity ions can be generated based on a variety of volumetric distributions, from a uniform random to more clustered ones, based on a correlation length parameter $\lambda$. For a non-zero $\lambda$, the number of impurity clusters, $N_\mathrm{c}$, is calculated by $N_\mathrm{c} = A/\lambda^2$, where $A$ is the two-dimensional area of the graphene layer. The size (or diameter) of each individual cluster is picked from a uniform random distribution between $\lambda/3$ and $2\lambda/3$, so the average cluster size is $\lambda/2$. To initialize the impurities, we first distribute the position of the centers of the $N_\mathrm{c}$ clusters stochastically and then distribute individual impurity ions around these centers with a Gaussian distribution, whose mean is the cluster center and the standard deviation equals half of the individual cluster size. This procedure results in an overall distribution that has a spatial autocorrelation function (SACF) very close a Gaussian, $\exp(-r^2/\lambda^2)$, as shown in Fig. \ref{figimpdist}. $\lambda$ extracted from the Gaussian fit [Fig. \ref{figimpdist}(b)] and the impurity cluster size estimated directly from the full width at half maximum (FWHM) of the SACF of the impurity distribution are in good agreement. For $\lambda = 0$, we distribute all the impurity ions stochastically, obtaining a uniform random distribution.

\subsubsection{Assignment}
In order to calculate the electric field that results from the charge distribution in the domain, the charges first have to be assigned to the grid. This is done using the cloud-in-cell (CIC) charge assignment scheme, in which the charges are represented by finite-volume charge clouds \cite{541446}. The CIC scheme results in a smoother field distribution than the commonly used nearest-grid-point scheme, where, as the name indicates, the total charge of each particle is assigned to the nearest grid point. In the CIC scheme, for each particle in a given grid cell, a portion of the particle's charge is assigned to each one of the cell's $8$ grid points. The fraction of the charge, or weight, of a particle located at $(x,y,z)$ on the $n$-th grid point with position $(x_\mathrm{n},y_\mathrm{n},z_\mathrm{n})$ is given by
\begin{equation}
w_\mathrm{n}=\left(1-\frac{|x_\mathrm{n}-x|}{\Delta x}\right)\left(1-\frac{|y_\mathrm{n}-y|}{\Delta y}\right)\left(1-\frac{|z_\mathrm{n}-z|}{\Delta z}\right), \label{eqnweights}
\end{equation}
where $\Delta x$, $\Delta y$, and $\Delta z$ are the grid cell dimensions along $x$, $y$, and $z$. For carriers in graphene, the weights are non-zero only in the plane of the sheet, since their motion is confined to that plane.

\subsection{Field initialization}
\label{subsec2:2}
An initial electric field distribution satisfying Gauss's law can be calculated as the gradient of the electrostatic potential $\Phi$, which is obtained by solving Poisson's equation for the initial charge distribution. The initial charge distribution $\rho(i,j,k)$ at grid point $(i,j,k)$ is given by the sum of the weights of all the charges in the cells surrounding that point,
\begin{equation}
\rho(i,j,k)=\sum^N_{\mathrm{c}=1}\frac{q_\mathrm{c}w_\mathrm{c}(i,j,k)}{\Delta x\Delta y\Delta z},
\end{equation}
where $N$ represents the total number of charges in the grid cells surrounding $(i,j,k)$, $q_\mathrm{c}$ is the charge of particle $c$, and $w_\mathrm{c}(i,j,k)$ is the weight of particle $c$ at  point $(i,j,k)$, given by Eq.~(\ref{eqnweights}). Using $\rho(i,j,k)$, we solve Poisson's equation with the successive-over-relaxation (SOR) method \cite{Press:1989fk} to get the electrostatic potential $\Phi(i,j,k)$. We use periodic boundary conditions at the four bounding planes perpendicular to the graphene layer and the Dirichlet boundary condition with a vanishing potential on the top and bottom planes. The initial electric field distribution is then calculated from
\begin{subequations}
\begin{align}
E_\mathrm{x}\left(i+\frac{1}{2},j,k\right) &= -\left[\Phi(i+1,j,k)-\Phi(i,j,k)\right]/\Delta x ,\\
E_\mathrm{y}\left(i,j+\frac{1}{2},k\right) &= -\left[\Phi(i,j+1,k)-\Phi(i,j,k)\right]/\Delta y ,\\
E_\mathrm{z}\left(i,j,k+\frac{1}{2}\right) &= -\left[\Phi(i,j,k+1)-\Phi(i,j,k)\right]/\Delta z .
\end{align}
\end{subequations}

\begin{figure*}
\centering
\includegraphics[width=0.8\textwidth]{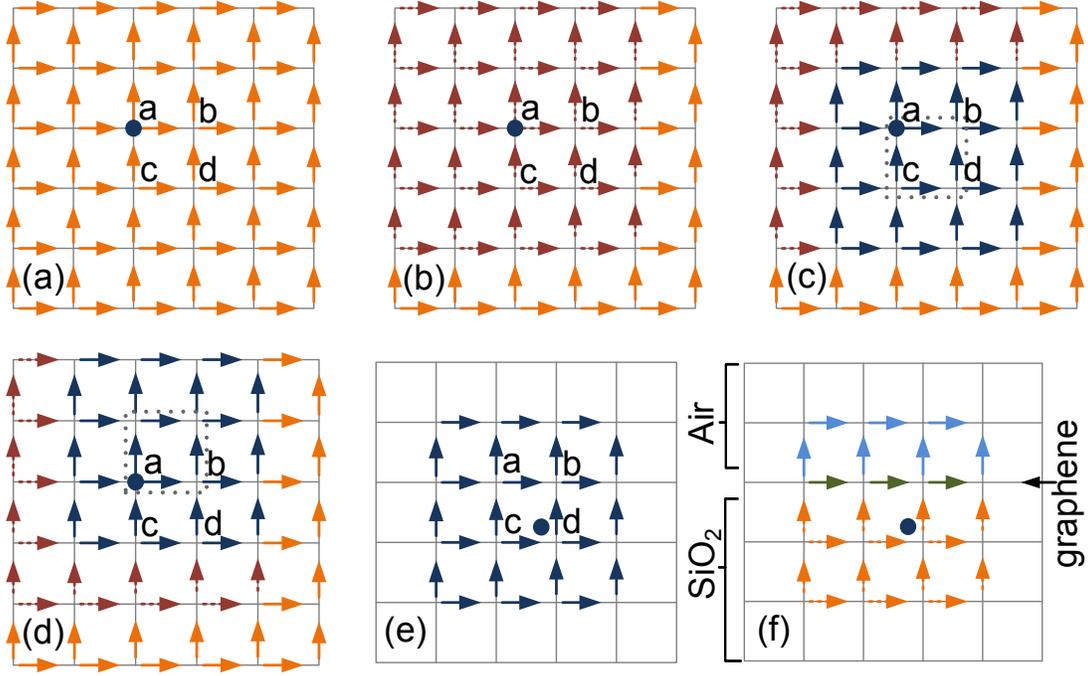}
\caption{(a)--(e) Illustration of the calculation of the grid-based electric field from an impurity ion. (a) A single ion (dark blue circle) placed at one of the corners of the cell $abcd$ and the corresponding electric field, marked by orange arrows, as calculated from the solution to Poisson's equation. (b) A smaller array of field values in the grid cells close to point $a$, marked by brown arrows, used for subsequent calculations. (c) The local fields (marked by dark blue arrows) due to a single ion at point $a$ found by properly shifting the stored fields around appropriate grid points. (d) The local fields due to a single ion at point $c$ using the solution for the ion at point $a$. (e) Local fields for a single charge located at an arbitrary position within the grid cell $abcd$, calculated using a weighted sum of the shifted potentials that correspond to a single ion at each of the corners. The weights are given by Eq.~\ref{eqnweights}. (f) Side view of the local 3D grid-based fields due to an impurity ion in the substrate near the interface of graphene, calculated as the weighted sum of the shifted fields in air (blue), graphene (green), and the $\rm{SiO_2}$ substrate (orange). Poisson's equation for a single ion is solved three times, for calculating fields due to an ion close to the interface, in air, graphene, and $\rm{SiO_2}$. The weighted sum is calculated based on the combination of fields in the appropriate medium for each impurity ion.}
\label{figDCfieldinit}
\end{figure*}

\noindent In addition to the initial electric field distribution, we also need the grid-based electric field contribution of the impurity ions, which is subtracted from the MD fields in the vicinity of the ions to avoid double counting \cite{willis:063714}. Since the ions are fixed in space, the grid-based contribution from the ions does not change in time; therefore we only have to calculate it once before commencing with the time-stepping. We rely on the linearity of Poisson's equation and simply use the solution for a single ion instead of solving the equation for each ion. Moreover, we use the MD contribution to the force on a carrier only for charges within a $3\times 3\times 3$-cell volume surrounding the given carrier and therefore only ions near the interface that are present within this cell volume are treated with MD. This approach is illustrated for a 2D grid in Fig.~\ref{figDCfieldinit}(a)--(e) and described  in further detail below.

We start by placing a single ion at grid point $a$ near the center of the domain, solve Poisson's equation and calculate the electric field, shown in Fig.~\ref{figDCfieldinit}(a) with orange arrows. From the complete field solution, we store the field values in the vicinity of cell $abcd$, marked in Fig.~\ref{figDCfieldinit}(b)--(d) by the brown arrows. With these stored fields, we can determine the short-range grid-based fields for an ion at points $a$, $b$, $c$ and $d$ simply by correctly shifting the stored fields to different points on the grid. For example, as shown in Fig.~\ref{figDCfieldinit}(c), marked by dark blue arrows are local fields for a single ion located at grid point $a$. For the purpose of illustration, the grey dashed box represents a secondary cell and marks the relative position of the ion. Now, by moving the grey dashed box to the cell above $abcd$, the position of the ion (at point $a$) relative to the grey box is equivalent to that of grid point $c$ relative to the cell $abcd$. Thus, using the original position of the ion (at grid point $a$), the local fields due to an ion at grid point $c$ (marked by dark blue arrows) can be found simply by shifting the grey dashed box, as shown in Fig.~\ref{figDCfieldinit}(d). Once, the local fields due to a single ion at all the corners of the grid cell $abcd$ are known, the field due to an ion at any arbitrary position within cell $abcd$ [Fig.~\ref{figDCfieldinit}(e)] can be calculated as a weighted sum of these shifted fields, where the weights are given by Eq.~(\ref{eqnweights}) for that impurity ion.

This procedure of storing and shifting short-range gird-based fields \cite{willis:063714} is applicable only within a single uniform medium. However, in order to have correct continuity in the fields near the interface of air, graphene and $\rm{SiO_2}$, the potentials in the respective mediums are required. A discontinuity in the fields at the interface results in residual fields that produce unphysical $dc$ current components that persist even without any externally applied field. Therefore we solve Poisson's equation three times -- in air, graphene, and $\mathrm{SiO_2}$ -- for the impurity ions near the interface of graphene and the substrate. The grid-based electric field contributions of the ions near the interface are then calculated by taking a combination of the appropriate fields from the three mediums, as shown in Fig.~\ref{figDCfieldinit}(f).


Calculating the grid-based field contributions from carriers using the above method is more complicated than for ions because of carrier motion. An implementation similar to the ions would be computationally burdensome as it would require recalculating the weights and shifting of the stored local fields at each time step. Therefore, we use the corrected-Coulomb scheme \cite{willis:063714} for determining the local grid-based field contributions of the carriers. In this scheme, a carrier is placed at a fixed location at a grid point. The grid-based field contribution is found from the solution to the Poisson's equation. The same Poisson's solution is used to determine the fields when the carrier is displaced by a small amount within the grid cell using the new weights of the carriers. The corrected-Coulomb fields are then calculated by subtracting these grid-based fields from the MD fields at the carrier locations. Although this method is not exact, the reduction in computational burden is significant and errors introduced have very little impact on the results \cite{willis:063714}.

\subsection{Current density calculation}
\label{subsec2:3}
At the initialization stage, Poisson's equation is solved to obtain the electric field profile consistent with the initial charge distribution. Thereafter, the continuity equation needs to enforced at each time step because FDTD and the continuity equation together ensure that Gauss's law remain satisfied throughout the simulation \cite{willis:063714,taflove2005computational}. We achieve this here by using the Villasenor-Buneman method \cite{Villasenor1992306} to calculate the current density, $\vec{J}$, from the change in the carrier position over a time step and assign it to the grid using the same CIC scheme as before. The current densities are defined at the same locations on the grid as the corresponding fields (see Fig. \ref{figfdtdgrid}) and are given by \cite{anuayw}
\begin{subequations}\label{eqncurrentdensity}
\begin{align}
J_\mathrm{x}\left(i+\frac{1}{2},j,k\right) &= \sum_{\mathrm{c}=1}^N\frac{q_\mathrm{c}}{\Delta x\Delta yt_\mathrm{g}}\frac{x_\mathrm{f}-x_\mathrm{i}}{\Delta t}\left(1-\frac{y_\mathrm{f}+y_\mathrm{i}}{2\Delta y}+j\right), \\
J_\mathrm{y}\left(i,j+\frac{1}{2},k\right) &= \sum_{\mathrm{c}=1}^N\frac{q_\mathrm{c}}{\Delta x\Delta yt_\mathrm{g}}\frac{y_\mathrm{f}-y_\mathrm{i}}{\Delta t}\left(1-\frac{x_\mathrm{f}+x_\mathrm{i}}{2\Delta x}+i\right),
\end{align}
\end{subequations}
where subscripts $f$ and $i$ represent the final and initial positions, respectively, and $t_\mathrm{g}$ is the thickness of graphene. Although the carriers move in a 2D plane, the current density must have the units of $\ampere\meter^{-3}$ for the sourcing of FDTD fields. Therefore, to calculate the current density in the correct units, we divide by $t_\mathrm{g}$. We use $t_\mathrm{g}\approx6\usk\angstrom$ to represent the approximate thickness of the graphene layer \cite{Lee18072008,4137639}. Motion of carriers into the neighboring grid cell is treated by dividing the path into sections, such that the motion in each cell is treated individually.

\subsection{Lorentz force calculation}
\label{subsec1:4}
We require the fields at the location of the carriers to calculate the force acting on the carriers drifting in EMC, Eq.~(\ref{eqnLorentzforce}). To determine the fields at the location of a carrier that is found inside a given grid cell, for each of the 8 grid points of that cell we first average the fields on the grid faces and grid lines (Fig. \ref{figfdtdgrid}) surrounding it. From the fields at each grid point, we can use the same CIC weights of each carrier to interpolate the fields and obtain the values at the carrier's actual location. The total electric field $E^\mathrm{c}_\mathrm{T}$ that accelerates carrier $c$ in EMC thus consists of the following contributions:
\begin{eqnarray}
E^\mathrm{c}_\mathrm{T}&=\sum^8_{\mathrm{n}=1}E^\mathrm{n}_{\mathrm{FDTD}}w^\mathrm{c}_\mathrm{n}+\sum_{\substack{\mathrm{c}'=1\\ \mathrm{c'\neq c}}}^N \left(E_{\mathrm{MD}}^{\mathrm{c\text{-}c'}}-E^{\mathrm{c}'}_{\mathrm{grid}}\right) \nonumber\\
&+ \sum_{\mathrm{i}=1}^M \left(E_{\mathrm{MD}}^{\mathrm{c\text{-}i}}-E^{\mathrm{i}}_{\mathrm{grid}}\right),
\end{eqnarray}
where $n$ enumerates the $8$ corners of a grid cell containing carrier $c$, and $w_\mathrm{n}$ are the CIC weights. $N$ and $M$ are the total numbers of carriers and ions, respectively, within the $27$ grid cells surrounding $c$. $E^\mathrm{n}_{\mathrm{FDTD}}$ is the FDTD field contribution, $E_{\mathrm{MD}}^{\mathrm{c\text{-}c'}}$ and $E_{\mathrm{MD}}^{\mathrm{c\text{-}i}}$ are the MD field contributions due to carrier-carrier and carrier-ion interactions respectively, while $E^{\mathrm{c'}}_{\mathrm{grid}}$, $E^{\mathrm{i}}_{\mathrm{grid}}$ are the local grid-based field contributions due to carriers and ions, respectively. The MD fields are pre-calculated before starting the time-stepping loop for a dense mesh in the real space and $k$-space within a volume of $3\times3\times3$ grid cells and stored in look-up tables. At any given time step, we then look up the MD fields based on the pairwise differences between carrier positions.

\section{Example: Conductivity of Graphene}
\label{sec:4}
In this section, we use the coupled EMC/FDTD/MD solver to calculate the \emph{dc} and \emph{ac} conductivity of graphene. The complex conductivity is computed from the spatially averaged values of the current density $\hat{J}(\omega)$ and electric field $\hat{E}(\omega)$ phasors as
\begin{equation}\label{eqnSigmaPhasors}
\sigma(\omega)=\frac{\hat{E}(\omega)\cdot\hat{J}^*(\omega)}{|\hat{E}(\omega)|^2}.
\end{equation}
The phasor quantities are calculated at each grid point in the graphene plane by using on-the-fly discrete Fourier transform of the time-dependent vector components after a steady state has been reached, then spatially averaging the components for use in (\ref{eqnSigmaPhasors}). For example, the current density phasor is given by
\begin{equation}\label{eqnsigma}
\hat{J}(\omega)=\sum^{T_s}_{n=t_s}\vec{J}\left[\cos{(2\pi f_0n\Delta t)}-i\sin{(2\pi f_0n\Delta t)}\right],
\end{equation}
where $t_s$ is the time to reach a steady state, $T_s$ is the total simulation time, $f_0$ is the frequency of external excitation ($f_0=0$ for \emph{dc} excitation), and $\vec{J}=J_\mathrm{x}\vec{x}+J_\mathrm{y}\vec{y}$ is the current density calculated from Eq.~(\ref{eqncurrentdensity}). Being that graphene is a 2D material, its conductivity is typically measured and presented in the units of $e^2/h$, the quantum of conductance. In order to convert the conductivity calculated in Eq.~(\ref{eqnsigma}) to those units, we multiply by a factor of $t_\mathrm{g}h/e^2$, where, as before, $t_g\approx 6$ {\AA} is the effective thickness of the graphene electron system \cite{Lee18072008,4137639}.

\subsection{\emph{dc} Conductivity}
\label{subsec4:1}
\begin{figure}
\centering
\includegraphics[width=\columnwidth]{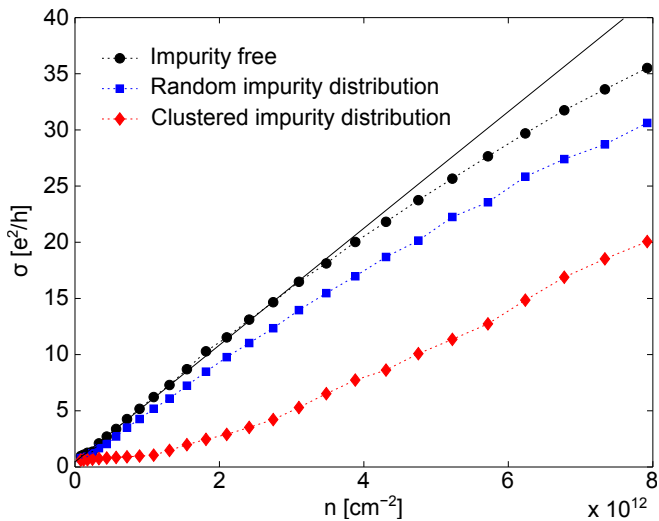}
\caption{\emph{dc} conductivity of supported graphene as a function of the carrier density. Black circles denote the results for impurity-free case, blue diamonds for a uniform random distribution,  and red diamonds for a clustered distribution ($40\usk\nano\meter$ average cluster size) of charged impurities with a sheet density of $5\times10^{11}\usk\centi\meter^{-2}$. The black line is a linear fit to the low-density ($<3\times10^{12}\usk\centi\meter^{-2}$) part of the impurity-free curve.}
\label{figdcsigma}
\end{figure}

We calculate the \emph{dc} conductivity of graphene as a function of the carrier sheet density, shown in Fig.~\ref{figdcsigma}, for an impurity sheet density of $5\times10^{11}\usk\centi\meter^{-2}$ with a uniform random distribution (blue squares) and a clustered distribution (red diamonds; correlation length of $40\usk\nano\meter$) and compare it with the conductivity of impurity-free graphene (black circles). These results reproduce  important features of the conductivity vs. carrier density curve observed in experiment \cite{Chen:2008fk}. The curve displays a sublinear increase at high carrier densities ($>4\times10^{12}\usk\centi\meter^{-2}$) for ``clean'' graphene. Moreover, for a given sheet density of charged impurities, the clustered impurity distribution results in lower conductivity than the uniform random one. This behavior has also been observed in experiment \cite{PhysRevLett.107.206601} and predicted in the calculations of carrier-impurity scattering rates with a structure factor describing correlations \cite{PhysRevLett.107.156601}. Our EMC/FDTD/MD simulation makes no assumptions about the screening length or structure factor, and uses real-space impurity positions and the corresponding carrier-impurity interactions to calculate the conductivity. Our results also show a flattening of the conductivity curve near the Dirac point for clustered impurity distributions, similar to that observed in conductivity measurements involving intentional potassium doping \cite{Chen:2008fk}.

\subsection{\emph{ac} Conductivity}
\label{subsec4:2}
\begin{figure}
\centering
\includegraphics[width=\columnwidth]{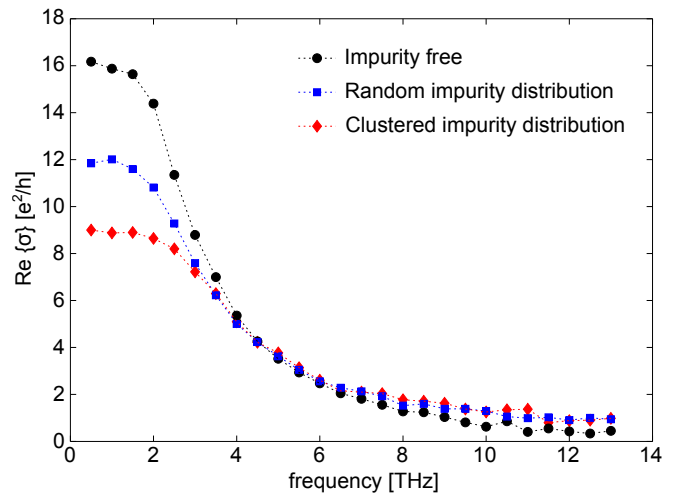}
\caption{Frequency-dependent \emph{ac} conductivity of supported graphene for the same charged impurity distribution as in Fig.~\ref{figdcsigma}. Carrier density is assumed to be $3\times10^{12}\usk\centi\meter^{-2}$.}
\label{figacsigma}
\end{figure}

The frequency-dependent \emph{ac} conductivity, shown in Fig.~\ref{figacsigma}, is calculated for the same impurity density and distributions as the \emph{dc} case (Fig.~\ref{figdcsigma}). Here we use a carrier density of $3\times10^{12}\usk\centi\meter^{-2}$. The frequency of the external excitation is varied from $500\usk\giga\hertz$ to $13\usk\tera\hertz$. In this range, carrier transport is dominated by intraband processes \cite{choi:172102} and is captured very well in our simulation. These results are in line with experimental measurements of frequency-dependent conductivity \cite{PhysRevB.83.165113,choi:172102}. For frequencies greater than $4\usk\tera\hertz$, our results show that the total impurity density and distribution do not affect the conductivity of graphene. However, for lower frequencies ($<4\usk\tera\hertz$), there is a significant dependence of conductivity on the impurity density and distribution. (As expected, the low-frequency conductivity limit obtained from \emph{ac} calculations is very close to the values calculated in the \emph{dc} simulations.)

\section{Summary}
\label{sec:5}
We have presented the implementation and application of the coupled EMC/FDTD/MD simulation technique to carrier transport in supported graphene in the presence of charged impurities. We have described the constituent techniques, as well as the important steps for their self-consistent coupling, such as charge initialization and assignment to the grid, field initialization, current density calculation based on particle motion in the gird, and avoiding double counting of the fields from FDTD and MD. The general implementation can also be applied to transport simulations of other 2D or quasi-2D materials.

We have demonstrated the use of the EMC/FDTD/MD method by calculating the \emph{dc} and \emph{ac} conductivity of supported graphene. The calculated \emph{dc} conductivity as a function of the carrier density reproduces the important features observed in experiments, such as the sublinear increase at high carrier density in clean samples \cite{Chen:2008fk} and flattening of the curve near the Dirac point for clustered impurity distribution \cite{PhysRevLett.107.206601}. The calculated \emph{ac} conductivity agrees with experimental observations \cite{PhysRevB.83.165113,choi:172102}.


\bibliographystyle{spphys}       

\begin{thebibliography}{10}
\providecommand{\url}[1]{{#1}}
\providecommand{\urlprefix}{URL }
\expandafter\ifx\csname urlstyle\endcsname\relax
  \providecommand{\doi}[1]{DOI \discretionary{}{}{}#1}\else
  \providecommand{\doi}{DOI \discretionary{}{}{}\begingroup
  \urlstyle{rm}\Url}\fi

\bibitem{Das-Sarma:2011fk}
S.~Das~Sarma, S.~Adam, E.H. Hwang, E.~Rossi, Rev. Mod. Phys. \textbf{83}, 407
  (2011)

\bibitem{Fontana:2013uq}
M.~Fontana, T.~Deppe, A.K. Boyd, M.~Rinzan, A.Y. Liu, M.~Paranjape, P.~Barbara,
  Sci. Rep. \textbf{3}(1634) (2013)

\bibitem{Zhang:2006uq}
P.~Zhang, E.~Tevaarwerk, B.N. Park, D.E. Savage, G.K. Celler, I.~Knezevic, P.G.
  Evans, M.A. Eriksson, M.G. Lagally, Nature \textbf{439}, 703 (2006)

\bibitem{Lin05022010}
Y.M. Lin, C.~Dimitrakopoulos, K.A. Jenkins, D.B. Farmer, H.Y. Chiu, A.~Grill,
  P.~Avouris, Science \textbf{327}(5966), 662 (2010)

\bibitem{B.:2011vn}
R.~B., R.~A., B.~J., G.~V., K.~A., Nat. Nano. \textbf{6}, 147 (2011)

\bibitem{Bonaccorso:2010ys}
F.~Bonaccorso, Z.~Sun, T.~Hasan, A.C. Ferrari, Nat. Photon. \textbf{4}, 611
  (2010)

\bibitem{doi:10.1021/nn2024557}
Z.~Yin, H.~Li, H.~Li, L.~Jiang, Y.~Shi, Y.~Sun, G.~Lu, Q.~Zhang, X.~Chen,
  H.~Zhang, ACS Nano \textbf{6}(1), 74 (2012)

\bibitem{Ju:2011kx}
L.~Ju, B.~Geng, J.~Horng, C.~Girit, M.~Martin, Z.~Hao, H.A. Bechtel, X.~Liang,
  A.~Zettl, Y.R. Shen, F.~Wang, Nat. Nano. \textbf{6}, 630 (2011)

\bibitem{C1CS15270J}
Y.~Liu, X.~Dong, P.~Chen, Chem. Soc. Rev. \textbf{41}, 2283 (2012)

\bibitem{sonde:132101}
S.~Sonde, F.~Giannazzo, C.~Vecchio, R.~Yakimova, E.~Rimini, V.~Raineri, Appl.
  Phys. Lett. \textbf{97}(13), 132101 (2010)

\bibitem{PhysRevB.77.195415}
S.~Fratini, F.~Guinea, Phys. Rev. B \textbf{77}, 195415 (2008)

\bibitem{Adam20112007}
S.~Adam, E.H. Hwang, V.M. Galitski, S.~Das~Sarma, P. Natl. Acad. Sci. USA
  \textbf{104}(47), 18392 (2007)

\bibitem{PhysRevLett.98.186806}
E.H. Hwang, S.~Adam, S.~Das~Sarma, Phys. Rev. Lett. \textbf{98}, 186806 (2007)

\bibitem{willis:063714}
K.J. Willis, S.C. Hagness, I.~Knezevic, J. Appl. Phys. \textbf{110}(6), 063714
  (2011)

\bibitem{tomizawa1993numerical}
K.~Tomizawa, \emph{Numerical Simulation of Submicron Semiconductor Devices}.
\newblock Electronic Materials and Devices Library (Artech House, 1993)

\bibitem{jacoboni1989monte}
C.~Jacoboni, P.~Lugli, \emph{The Monte Carlo Method for Semiconductor Device
  Simulation}.
\newblock Computational Microelectronics (Springer, 1989)

\bibitem{xaldgh}
J.~Ayubi-Moak, S.~Goodnick, S.~Aboud, M.~Saraniti, S.~El-Ghazaly, J. Comput.
  Elec. \textbf{2}(2-4), 183 (2003)

\bibitem{zskugadf}
J.~Ayubi-Moak, S.~Goodnick, M.~Saraniti, J. Comput. Elec. \textbf{5}(4), 415
  (2006)

\bibitem{anuayw}
K.~Willis, J.~Ayubi-Moak, S.~Hagness, I.~Knezevic, J. Comput. Elec.
  \textbf{8}(2), 153 (2009)

\bibitem{willis:062106}
K.J. Willis, S.C. Hagness, I.~Knezevic, Appl. Phys. Lett. \textbf{96}(6),
  062106 (2010)

\bibitem{PhysRevLett.56.1295}
P.~Lugli, D.K. Ferry, Phys. Rev. Lett. \textbf{56}, 1295 (1986)

\bibitem{Ferry1991119}
D.K. Ferry, A.M. Kriman, M.J. Kann, R.P. Joshi, Comput. Phys. Commun.
  \textbf{67}(1), 119  (1991)

\bibitem{822288}
C.~Wordelman, U.~Ravaioli, Electron Devices, IEEE Transactions on
  \textbf{47}(2), 410 (2000)

\bibitem{Vasileska:2008:1546-1955:1793}
D.~Vasileska, H.~Khan, S.~Ahmed, J. Comput. Theor. Nanosci. \textbf{5}(9), 1793
  (2008)

\bibitem{willis:122113}
K.J. Willis, S.C. Hagness, I.~Knezevic, Appl. Phys. Lett. \textbf{102}(12),
  122113 (2013)

\bibitem{PhysRevLett.107.156601}
Q.~Li, E.H. Hwang, E.~Rossi, S.~Das~Sarma, Phys. Rev. Lett. \textbf{107},
  156601 (2011)

\bibitem{RevModPhys.55.645}
C.~Jacoboni, L.~Reggiani, Rev. Mod. Phys. \textbf{55}, 645 (1983)

\bibitem{sule:053702}
N.~Sule, I.~Knezevic, J. Appl. Phys. \textbf{112}(5), 053702 (2012)

\bibitem{PhysRevB.81.121412}
K.M. Borysenko, J.T. Mullen, E.A. Barry, S.~Paul, Y.G. Semenov, J.M. Zavada,
  M.B. Nardelli, K.W. Kim, Phys. Rev. B \textbf{81}, 121412 (2010)

\bibitem{PhysRevB.82.115452}
A.~Konar, T.~Fang, D.~Jena, Phys. Rev. B \textbf{82}, 115452 (2010)

\bibitem{taflove2005computational}
A.~Taflove, S.~Hagness, \emph{Computational Electrodynamics: The
  Finite-Difference Time-Domain Method}.
\newblock The Artech House antenna and propagation library (Artech House,
  Incorporated, 2005)

\bibitem{1138693}
K.~Yee, IEEE T. Antenn. Propag. \textbf{14}(3), 302 (1966)

\bibitem{MOP:MOP14}
J.A. Roden, S.D. Gedney, Microw. Opt. Techn. Lett. \textbf{27}(5), 334 (2000)

\bibitem{rapaport2004art}
D.~Rapaport, \emph{The Art of Molecular Dynamics Simulation} (Cambridge
  University Press, 2004)

\bibitem{alder:459}
B.J. Alder, T.E. Wainwright, J. Chem. Phys. \textbf{31}(2), 459 (1959).
\newblock \doi{10.1063/1.1730376}

\bibitem{PhysRevLett.65.1619}
A.M. Kriman, M.J. Kann, D.K. Ferry, R.~Joshi, Phys. Rev. Lett. \textbf{65},
  1619 (1990)

\bibitem{joshi:2369}
R.P. Joshi, A.M. Kriman, M.J. Kann, D.K. Ferry, Appl. Phys. Lett.
  \textbf{58}(21), 2369 (1991)

\bibitem{PhysRevB.61.7353}
P.~Gori-Giorgi, F.~Sacchetti, G.B. Bachelet, Phys. Rev. B \textbf{61}, 7353
  (2000)

\bibitem{LiangACSNano}
X.~Liang, B.A. Sperling, I.~Calizo, G.~Cheng, C.A. Hacker, Q.~Zhang, Y.~Obeng,
  K.~Yan, H.~Peng, Q.~Li, X.~Zhu, H.~Yuan, A.R. Hight~Walker, Z.~Liu, L.m.
  Peng, C.A. Richter, ACS Nano \textbf{5}(11), 9144 (2011)

\bibitem{fang:092109}
T.~Fang, A.~Konar, H.~Xing, D.~Jena, Appl. Phys. Lett. \textbf{91}(9), 092109
  (2007)

\bibitem{541446}
S.~Laux, IEEE Trans. Comput-Aided Des. Integr. Circuits Syst. \textbf{15}(10),
  1266 (1996)

\bibitem{Press:1989fk}
W.H. Press, \emph{Numerical Recipes : The Art of Scientific Computing}
  (Cambridge University Press, 1989)

\bibitem{Villasenor1992306}
J.~Villasenor, O.~Buneman, Comput. Phys. Commun. \textbf{69}(2--3), 306  (1992)

\bibitem{Lee18072008}
C.~Lee, X.~Wei, J.W. Kysar, J.~Hone, Science \textbf{321}(5887), 385 (2008)

\bibitem{4137639}
M.C. Lemme, T.~Echtermeyer, M.~Baus, H.~Kurz, IEEE Electr. Device L.
  \textbf{28}(4), 282 (2007)

\bibitem{Chen:2008fk}
J.H. Chen, C.~Jang, S.~Adam, M.S. Fuhrer, E.D. Williams, M.~Ishigami, Nat.
  Phys. \textbf{4}(5), 377 (2008)

\bibitem{PhysRevLett.107.206601}
J.~Yan, M.S. Fuhrer, Phys. Rev. Lett. \textbf{107}, 206601 (2011)

\bibitem{choi:172102}
H.~Choi, F.~Borondics, D.A. Siegel, S.Y. Zhou, M.C. Martin, A.~Lanzara, R.A.
  Kaindl, Applied Physics Letters \textbf{94}(17), 172102 (2009)

\bibitem{PhysRevB.83.165113}
J.~Horng, C.F. Chen, B.~Geng, C.~Girit, Y.~Zhang, Z.~Hao, H.A. Bechtel,
  M.~Martin, A.~Zettl, M.F. Crommie, Y.R. Shen, F.~Wang, Phys. Rev. B
  \textbf{83}, 165113 (2011)

\end{thebibliography}

\end{document}